\documentclass[12pt]{article}
\usepackage{epsfig}

\def\beq{\begin{equation}}
\def\eeq#1{\label{#1}\end{equation}}
\def\eeqn{\end{equation}}

\def\beqa{\begin{eqnarray}}
\def\eeqa#1{\label{#1}\end{eqnarray}}
\def\eeqan{\end{eqnarray}}

\let\bar=\overbar

\def\Dslash{\not{\hbox{\kern-4pt $D$}}}
\def\dslash{\not{\hbox{\kern-2pt $\del$}}}

\def\msb{{\bar{\ssstyle M \kern -1pt S}}}

\def\Title#1{\begin{center} {\Large {\bf #1} } \end{center}}

\begin{document}

\Title{Precise determination of V$_{ud}$ and V$_{us}$}

\begin{center}
Summary of the WG1 contributions, \\
written for the\\
Proceedings of CKM 2012,\\
the 7th International Workshop on the CKM Unitarity Triangle,\\
University of Cincinnati, USA, 28 September - 2 October 2012      
\end{center}
\vspace{0.3cm}

\bigskip\bigskip


\begin{raggedright}  

{\it Takashi Kaneko, KEK and\\
Barbara Sciascia, INFN-LNF}
\bigskip\bigskip
\end{raggedright}

\begin{abstract}
The actual limit of the Unitarity condition of the first row of the CKM matrix 
$|V_{ud}|^2+|V_{us}|^2+|V_{ub}|^2=1+\Delta_{CKM}$ is $\Delta_{CKM}=-0.0001(6)$.
In 2010 the same was $\Delta_{CKM,2010}=+0.0001(6)$.
Despite the only difference of a sign, and with an absolute change of the value of one third of the accuracy,
a substantial amount of work has been done in the last two years to improve the knowledge of all the contributions
to this stringent limit to CKM unitarity, and more is expected in the next years.
In this paper we present an organized summary of all the important contributions presented during the WG1 sessions,
referring as much as possible to the contribution papers prepared by the individual authors.
\end{abstract}

New bounds on violations of CKM unitarity translate into significant constraints on various 
new physics scenarios. Such tests may eventually turn up evidence of new physics. 
If the couplings of the W to quarks and leptons are indeed specified by a single
gauge coupling, then for universality to be observed as the equivalence of the Fermi
constant $G_F$ as measured in muon and hadron decays, the CKM matrix must be unitary. 
Currently, the most stringent test of CKM unitarity is obtained from the
first-row relation $|V_{ud}|^2+|V_{us}|^2+|V_{ub}|^2=1+\Delta_{CKM}$; here we
shortly discuss the ingredients that contribute to its accuracy and 
its possible new physics implications.

We start with Vincenzo Cirigliano~\cite{Cirigliano} that in his talk 
emphasized a mo\-del-in\-de\-pen\-dent EFT approach to $\beta$ decays and
Cabibbo universality tests. 
Given the hierarchy $|V_{ud}|^2 \gg |V_{us}|^2$, let us focus on 
effects of physics beyond the Standard Model (BSM) to $|V_{ud}|$.
Assuming that right-handed neutrinos do not appear as low-energy degrees of freedom,
new physics introduces five operators
\begin{eqnarray}
  \epsilon_\Gamma \, \bar{l}\gamma_\mu(1-\gamma_5)\nu_l \cdot \bar{u} \Gamma^\mu d
  \hspace{3mm}
  (\Gamma=V,A), 
  \hspace{10mm}
  \epsilon_\Gamma \, \bar{l}(1-\gamma_5)\nu_l \cdot \bar{u} \Gamma d
  \hspace{3mm}
  (\Gamma=S,P,T)
  \label{eqn:bsm_interaction:lhnu}
\end{eqnarray}
into the Lagrangian of the $d \to u l \nu_l$ transitions.
The (axial-)vector and scalar couplings $\epsilon_{\{V,A,S\}}$ are constrained 
at the level of $10^{-3}$ from the superallowed nuclear $\beta$ decays. 
The current precise knowledge of $|V_{ud}|$ from the nuclear decays 
together with a future competitive determination from neutron decays 
could constrain $\epsilon_{\{S,T\}}$ at 0.02\% level. 
The pseudo-scalar coupling $\epsilon_P$ is strongly constrained 
from the ratio $\Gamma(\pi \to e\nu_e)/\Gamma(\pi \to \mu\nu_\mu)$. 
The outlook is therefore quite positive: 
the effective couplings of all the BSM charged-current operators are currently probed 
or will be soon probed at the level of $10^{-3}$ or better. 
This corresponds to probing maximal BSM physics scales $\Lambda$ ranging from 7 TeV 
(for scalar and tensor interactions) to 11 TeV (for vector interactions). 

New physics can also modify the Lagrangian of the muon decays.
Its effects would be encoded in the Fermi constant $G_F$,
which is best determined by the measurement of the positive muon lifetime $\tau_\mu$.
Tim Gorringe~\cite{Gorringe} reported results from the MuLan measurement of the positive muon 
lifetime, conducted at the Paul Scherrer Institute.
The result is characterized by a part-per-million accuracy, largely dominated by the 
statistical contribution, and rock-solid systematic effects study.
The MuLan measurement translates into $G_F = 1.166 378 7(6)\times10^{-5} \mbox{GeV}^{-2}$.
The 0.5 ppm error is dominated by the 1.0 ppm uncertainty of the lifetime measurement, 
with contributions of 0.08 ppm from the muon mass measurement and 0.14 ppm from the theoretical 
corrections. $\tau_\mu$ and $G_F$ at 0.1 ppm is now the open challenge for the next years.

\vspace{0.2 cm}
$|V_{ud}|$ is best determined using $0^+\rightarrow 0^+$ super-allowed nuclear $\beta$ decays.
The $|V_{ud}|$ state of the art has been presented by Dan Melconian~\cite{Melconian}.
Given the limited number of new experimental contributions, no new survey on $|V_{ud}|$
has been done and its world average is unchanged from the 2010 CKM edition~\cite{CKM2010}. 
The  $0^+\rightarrow 0^+$ transitions benefit from the conservation
of the vector current and from small isospin-breaking corrections.
The experimental inputs are combined in a quantity that should be nucleus-independent to first order. 
This is true only applying also the isospin-breaking corrections, $\delta_C$. These are a dominant contribution
to the $|V_{ud}|$ accuracy and have been studied in recent years using a variety of
theoretical methods. A new approach, proposed by Towner and Hardy~\cite{Towner-Hardy}, allow one to
experimentally test the SU(2) correction. They succeeded in measuring the $\delta_C$
for the elements with the largest isospin-breaking correction, $^{32}Ar$ and $^{32}Cl$, 
and found it in agreement with the theoretical calculations.
Other groups are beginning to develop complementary models of SU(2) in superallowed decays
that will be important checks and may lead to smaller systematic error on this $|V_{ud}|$ determination.

\vspace{0.2 cm}
The alternative ways to determine $|V_{ud}|$, from neutron lifetime, from mirror decays, and
from pion $\beta$ decay, are still limited by the experimental accuracy.
It is nevertheless worth improving their accuracy because the involved processes probe different BSM operators.

As a probe free from nuclear structure corrections, the decay of the free neutron has the potential to provide
the most accurate value of $|V_{ud}|$. However, the experimental sensitivity still
needs to be further improved to become competitive with superallowed nuclear $\beta$ decays.
The current status of the neutron decay studies relevant for the determination of $|V_{ud}|$
has been presented by Oliver Zimmer~\cite{Zimmer}.

In the standard V-A theoretical description of neutron decay both the vector and the axial-vector
currents contribute, with $g_V$ and $g_A$ coupling constants respectively, so that two observables are 
needed to access $|V_{ud}|$: the neutron lifetime $\tau_n$ and the $g_A/g_V$ ratio.
For $\tau_n$, the accuracy needed to compete with $0^+\rightarrow 0^+$ transitions in the $|V_{ud}|$ determination
is $\sim$0.3 s on $\tau_n$ (a factor three from the present accuracy).
In this respect, a novelty is the solution of the 5.4 $\sigma$ tension between different determinations,
mainly due to a single 2005 measurement.
A wide effort in the field and a new measurement in 2010, pushed the authors of the 2005 result to
scrutinize their procedures and to recently publish a corrected value.
With these changes included, the tension is reduced to 1.4 $\sigma$ even if the the 1.8 scale
factor applied by the PDG indicates that systematic uncertainties are not properly
taken into account in all experiments.
The highest experimental sensitivity on the $g_A/g_V$ ratio has been achieved measuring 
the $\beta$ asymmetry coefficient. 
The accuracy goal here is $\sim$0.0003 on $g_A/g_V$, about one order of magnitude below 
the present determination.
For both $\tau_n$ and $g_A/g_V$ many projects are in the pipeline using ultra-cold neutrons or magnetic trapping.
Some are in a very advanced state and have the potential to reach the accuracy goal in the next years.

Nuclear mirror transitions occur between isobaric analogue states within an isospin
doublet, where initial and final states have the same spin and parity.
The determination of $|V_{ud}|$ from nuclear mirror transitions
and the current experimental efforts aimed at improving its precision
have been presented by Oscar Naviliat-Cuncic~\cite{Naviliat-Cuncic}.
In addition, the mirror transitions are driven by a mixing of vector and axial-vector interactions,
and the present $|V_{ud}|$ accuracy is dominated by the experimental error in the determination of the
mixing ratio.
As in the neutron lifetime case, two experimental inputs are needed. These are the half-life of the decay, and one of the following: 
the $\beta-\nu$ angular correlation, the $\beta$ asymmetry, or the $\nu$ asymmetry.
Recently a substantial activity has been initiated on the experimental and theoretical sides to improve all
the relevant inputs: many experiments are on going, applying different techniques, to measure 
the $\beta-\nu$ angular correlation, and there are plans to measure also  relative $\beta$  asymmetries.
These efforts should enable significant improvements in the precision on $|V_{ud}|$ from mirror transitions,
which is currently a factor $\sim$8 less precise than the value extracted from Fermi transitions.

\vspace{0.2 cm}
Moving to $|V_{us}|$, its best determination arises from $K_{\ell3}$ and $K_{\ell2}$  decays.
A very comprehensive review of the $|V_{us}|$ determination from kaons and of its effects on
the unitarity test of the first row has been presented by Matthew Moulson~\cite{Moulson}.
In the field, there have been
a few significant new measurements and some important theoretical developments.
The experimental inputs for the determination of $|V_{us}|$ from $K_{\ell3}$ decays are the
rates and form factors for the decays of both charged and neutral kaons. There have
been no new branching ratio measurements since the 2010 review. On the other
hand, both the KLOE and KTeV collaborations have new measurements of the $K_S$
lifetime. Finally, the NA48/2 experiment has recently released preliminary results for the form
factors for charged kaon decays.
This is important because it helps to resolve a controversy: the older measurements of 
the $K_{\mu3}$ form factors for $K_L$ decays from NA48 are in such strong disagreement with 
the other existing measurements that they have been excluded from the FlaviaNet averages. 
The new NA48/2 measurements, on the other hand, are in good agreement with other measurements.
For all of the above efforts, however, the value of and uncertainty on $|V_{us} f_+(0)|$ are
essentially unchanged. This is because the new results are nicely consistent with the
older averages, and neither the $K_S$ lifetime nor the phase space integrals were significant
contributors to the overall experimental uncertainty.
The latter is dominated by the lifetime accuracy for the $K_L$ and by the branching ratio
for the $K_S$ and $K^\pm$ determinations. In the near future there are no kaon experiments
planning new branching ratio or lifetime measurements.
For the $K^\pm$, also the uncertainty on the theoretical isospin-breaking correction gives the largest contributions
to the $|V_{us} f_+(0)|$ uncertainty.

Besides the latter, advances in algorithmic sophistication and computing power are
leading to more and better lattice QCD estimates of the hadronic constants $f_+(0)$
and $f_K/f_\pi$, which enter into the determination of $|V_{us}|$ from $K_{l3}$ and $K_{\mu2}$ decays,
respectively.
Due to their non-perturbative nature 
the only systematically improvable way to compute them are simulations of lattice QCD.
Since the ultimate goal is a test of the SM, any model-dependence should
be avoided and this is where progress in lattice simulations is currently being made.
In addition, two groups working on the classification and averaging of
results from lattice QCD have joined their efforts, constituting the newly formed
Flavor Lattice Average Group (FLAG-2)~\cite{FLAG2} to provide recommended values of these
constants.

The talk by Andreas J\"uttner~\cite{Juttner} reported on 
the status and ongoing improvements of determinations of $f_+(0)$. 
%
%
A key observation that allows a precise extraction of $f_+(0)$ is 
the conservation of the vector current at zero momentum transfer
in the $SU(3)$ limit: 
the normalization is fixed in this limit and 
corrections start at the second order in $SU(3)$ breaking effects. 
Because of this fortunate situation, 
recent lattice computations determine $f_+(0)$ at the level of 0.5\,--\,1.0\,\,\% 
and show an excellent agreement among them.
%
%
The uncertainty of $f_+(0)$ in the state-of-the-art calculations 
is dominated by the statistical error and 
the error due to the extrapolation to the physical pion mass. 
The latter is about to be removed by simulating very close to or at the physical mass.

In his talk Jack Laiho~\cite{Laiho} presented the lattice progress and the future prospects for $f_K/f_\pi$. 
This is a key input in the determination of $|V_{us}|/|V_{ud}|$ via $K_{l2}$ decays, 
which probe different BSM operators than the $K_{l3}$ decays. 
While the normalization in the $SU(3)$ limit is fixed,
only the axial-vector current contributes to the decay constants 
and $f_K/f_\pi$ can receives unsuppressed chiral corrections. 
However, recent simulations on realistic lattices, 
particularly those at small or even physical pion mass, 
have calculated this ratio to sub-percent precision 
leading to the world average with a 0.4\% accuracy. 
A further reduction of the discretization error as well as effects of finite lattice volumes 
is needed to improve the precision, say, to a level of 0.2\,\%.
We expect more simulations with different lattice formulations at the physical pion mass
in the future for better understanding and control of these systematic uncertainties. 


At the level of precision now reached in the lattice determinations of $f_+(0)$ and $f_K/f_\pi$,
the uncertainties of the electromagnetic and isospin corrections are becoming non-negligible. 
Traditionally, these corrections have been estimated by means of chiral perturbation theory (ChPT),
in which an extension to higher orders is generally not easy due to rapidly increasing numbers
of relevant diagrams and unknown effective couplings. 
An interesting possibility is to include these corrections into the lattice determinations 
in a fully non-perturbative way. 

In his talk Nazario Tantalo~\cite{Tantalo} presented a first principle lattice calculation of 
the QCD isospin corrections, namely those coming from the small difference of 
the up and down quark masses $\Delta m_{ud}\!=\!(m_u-m_d)/2$ 
in the absence of the electromagnetic interactions. 
Isospin is an approximate symmetry of QCD and most of the theoretical 
predictions on phenomenologically relevant
hadronic observables have been derived by assuming the exact validity of isospin
symmetry. This is also the case for most of the non-perturbative theoretical predictions 
on hadronic matrix elements obtained over the years by performing lattice QCD
simulations, like the hadronic constants $f_+(0)$ and $f_K/f_\pi$. 
The RM123 collaboration has recently performed a lattice calculation of the QCD isospin
corrections from numerical simulations of isosymmetric QCD combined with 
a systematic expansion of the partition function with respect to the small parameter 
$\Delta m_{ud}$. Their new method, therefore, does not need time-consuming simulations
of the isospin-broken theory. 
They obtained encouraging results for the isospin corrections, which are 
of the same order of magnitude, though higher, of the ChPT estimate. 
It should be noted that the separation of QED from QCD isospin-breaking effects 
is prescription-dependent. 
It is therefore important to calculate the full (QCD+QED) correction 
on the lattice, which could be an interesting alternative 
to the conventional ChPT approach.

\vspace{0.2cm}
The hadronic $\tau$ decays provide an alternative way to measure $|V_{us}|$ 
and to probe the relation of the first row of the CKM matrix.
Measurements of $|V_{us}|$ from $\tau$ decays are complementary to those from kaon decays  because
new physics scenarios that couple primarily to the third generation could cause deviation between measurements 
of $|V_{us}|$ in the kaon and $\tau$ systems.

In her talk Elvira G\'amiz~\cite{Gamiz} gave an overview of the theoretical issues related to 
the $|V_{us}|$ extractions from inclusive and exclusive hadronic $\tau$ decays.
Some exclusive decay channels, such as $\tau \!\to\! K\nu$ and $\tau \!\to\! K\pi\nu$,
can be used to extract $|V_{us}|$ in a similar manner to $K_{l2}$ and $K_{l3}$ decays.
This method is therefore sensitive to the same lattice QCD uncertainties.
The most precise measurement could be offered from the SU(3)-breaking effect in the inclusive rate 
\begin{eqnarray}
  \delta R
  & = & 
  \frac{R_{\tau,V+A}}{|V_{ud}|^2} - \frac{R_{\tau,S}}{|V_{us}|^2},
\end{eqnarray}
where 
$R_{\tau,S}\!=\!\Gamma(\tau \!\to\! X_s\nu_\tau)/\Gamma(\tau \to e\bar{\nu}_e\nu_\tau)$
is the Cabibbo-suppressed hadronic rate into strange particles ($X_s$)
and $R_{\tau,V+A}\!=\!\Gamma(\tau \!\to\! X_{{\rm non-}s}\nu_\tau)/\Gamma(\tau \to e\bar{\nu}_e\nu_\tau)$
is the Cabibbo-allowed hadronic rate. 
The SU(3)-breaking effect $\delta R$ can be estimated from 
finite energy sum rules (FESR). 
This technique makes the hadronic $\tau$ decays an ideal system to study low-energy QCD 
under rather clean conditions,
allowing the determination of the strong coupling with the same precision achieved by
lattice determinations. 



\begin{figure}[t]
\begin{center}
\epsfig{file=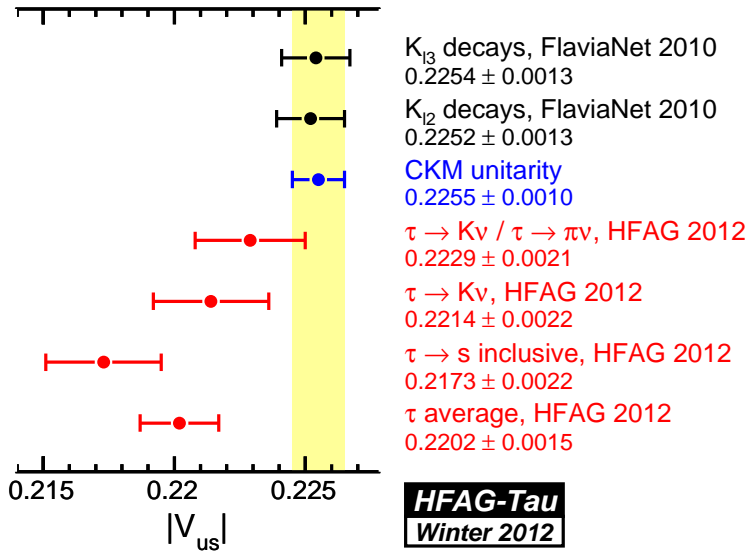,height=3.0in}
\caption{
  An update of $|V_{us}|$ from the hadronic $\tau$ decays
  (see Ref.~\cite{Nugent} and references therein).
  The three upper values are from the FlaviaNet Working Group's 
  analysis of the $K_{l3}$ and $K_{l2}$ decays in 2010
  and from the unitarity constraint and $|V_{ud}|$.
}
\label{fig:vus}
\end{center}
\end{figure}

To profit from $\delta R$, we need to measure the inclusive strange and non-strange spectral density
functions, which are constructed from the sum of invariant mass distributions for
each of the strange and non-strange decay modes and normalized to the corresponding
branching fractions. Since there are no solid predictions for
the branching fractions of hadronic individual $\tau$ decays, all possible modes must be
measured or upper bounds have to have placed on them.
This technique is completely independent of the kaon measurements, and if all of the branching fractions and spectral
functions were updated with the whole data sets of BELLE and BABAR experiments, this method would
be expected to provide the most precise measurement of $|V_{us}|$.

Ian Nugent~\cite{Nugent} presented an overview of the current status of the experiments. 
Currently, both the exclusive and the inclusive decays determine $|V_{us}|$ with an accuracy of about 1.0\,\%,
which is slightly larger than 0.6\,\% achieved in the determinations from $K_{l2}$ and $K_{l3}$ decays.
More importantly, 
the updated experimental results lead to only a slight change from the previous workshop~\cite{CKM2010}
in $|V_{us}|$ from the inclusive decays.
This result is about 3~$\sigma$ below the $K_{l2}$ and $K_{l3}$ determinations
as shown in Fig.~\ref{fig:vus}.

Since the uncertainty is limited by the experimental precision, 
further experimental data are needed before drawing any significant conclusion. 
While the reliability of the FESR analysis has been studied, 
a more thorough study, for instance about the stability against the choice of the weight, 
is also welcome.

\begin{figure}
\begin{center}
\includegraphics[height=0.35\textheight]{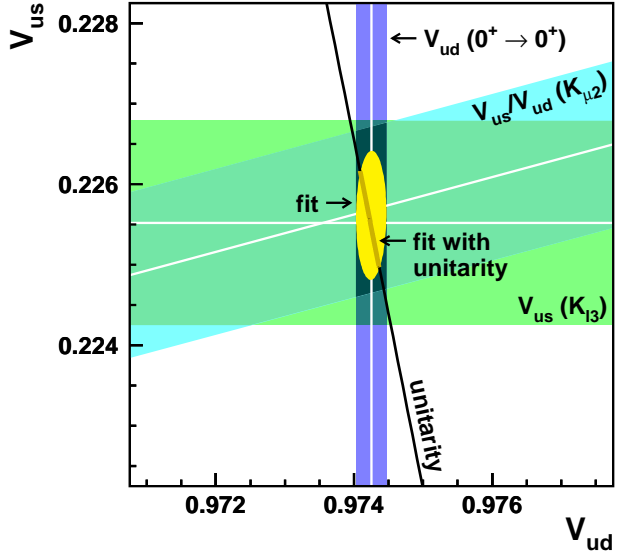}
\caption{$|V_{ud}|$ from $0^+\to0^+$ $\beta$ decays, $|V_{us}|$ from
$K_{l3}$ decays, and $|V_{us}/V_{ud}|$ from $K_{l2}$ decays
in the plane of ($|V_{ud}|$, $|V_{us}|$)~\cite{Moulson}.
The yellow ellipse indicates the $1\sigma$ confidence interval 
for the fit to fix $|V_{ud}|$ and $|V_{us}|$.
The unitarity constraint is shown by the solid line.
}
\label{fig:univers}
\end{center}
\end{figure}

\vspace{0.2 cm}
Presently, the most precise value of $|V_{ud}|$, 
\begin{eqnarray}
  |V_{ud}| = 0.97425(22){\mbox ,}
\end{eqnarray} 
is obtained from $0^+\rightarrow0^+$ nuclear decays and is unchanged from the previous CKM workshop.
This together with the updated values of $|V_{us}|\!=\!0.2254(13)$ from $K_{l3}$ decays and 
$|V_{us}/V_{ud}|\!=\!0.2317(11)$ from $K_{l2}$ can be combined in a single fit to determine 
the CKM elements and $\Delta_{CKM}$~\cite{Moulson}.
As plotted in Fig.~\ref{fig:univers},
the fit does not change the input value of $|V_{ud}|$ and yields
\begin{eqnarray}
  |V_{us}| = 0.2256(8),
  \hspace{10mm} 
  \Delta_{CKM} = + 0.0001(6)
\end{eqnarray}
in perfect agreement with unitarity.
The uncertainty of $\Delta_{CKM}$ is equally shared by $|V_{ud}|$ and $|V_{us}|$. 
The precision of $|V_{ud}|$ is determined from the uncertainty 
of the transition independent radiative correction~\cite{radcorr},
which has been stable over the last several years.
The uncertainty of $|V_{us}|$ is currently dominated by uncertainties
in the lattice results for $f_+(0)$ and $f_K/f_\pi$,  
which are at the level of 0.5\%. 
Thus, at the moment, the lattice offers the most certain prospects 
for further improvement.

\vspace{0.2 cm}
Although the accuracy of $\Delta_{CKM}$ is essentially unchanged from the previous workshop, 
the current precision already allow us to put significant constraints 
on new physics 
as summarized in the beginning of this paper. 
An interesting question is whether the low-energy observables 
from the nuclear, neutron, kaon and $\tau$ decays 
have a higher sensitivity than the rest of observables, 
for instance, from energy-frontier experiments. 
In his talk Mart\'in Gonz\'alez-Alonso~\cite{Gonzalez-Alonso} answered this question
by comparing new physics bounds from the low-energy observables with those obtained by the CMS Collaboration
analyzing 5 fb$^{-1}$ of data recorded at $\sqrt{s}=7$~TeV in the $pp \to e + MET$ (Missing Transverse Energy) channel.
Assuming that the heavy mediators that generate the BSM interactions 
in Eq.~(\ref{eqn:bsm_interaction:lhnu}) are too massive to be produced at the LHC,
we can again employ an EFT approach and put bounds on the $\epsilon_i$ couplings
from the LHC search. 
For the (axial) vector and pseudo-scalar ones low-energy probes are much more powerful.
There is an interesting competition between low- and high-energy searches
for the scalar and tensor couplings. 
In addition to Eq.~(\ref{eqn:bsm_interaction:lhnu}), 
we can also introduce BSM interactions involving right-hand neutrinos,
for which the LHC will dominate the search. 

Even if the LHC sensitivity will get better as more data is collected,
future low-energy experiments, such as with (ultra)cold neutrons, 
will improve in finding the bounds on new physics. 
Currently, the precise determination of $|V_{ud}|$ and $|V_{us}|$ provides 
one of the most stringent tests of the Standard Model and 
will play an important role, complementary to the collider searches, 
in probing new physics.

\end{document}